%% file: main.tex
\documentclass[iicol]{sn-jnl}

\usepackage{graphicx}%
\usepackage{multirow}%
\usepackage{amsmath,amssymb,amsfonts}%
\usepackage{amsthm}%
\usepackage{mathrsfs}%
\usepackage[title]{appendix}%
\usepackage{xcolor}%
\usepackage{textcomp}%
\usepackage{manyfoot}%
\usepackage{booktabs}%
\usepackage{algorithm}%
\usepackage{algorithmicx}%
\usepackage{algpseudocode}%
\usepackage{listings}%
\usepackage[separate-uncertainty=true]{siunitx}
\usepackage{verbatim}
\usepackage{placeins}
\usepackage{lineno}

\usepackage{orcidlink}
\input{NucleusAuthorlist}

\begin{document}

\let\cleardoublepage\clearpage

\title{Sub-keV Electron Recoil Calibration for Macroscopic Cryogenic Calorimeters using a Novel X-ray Fluorescence Source}

\abstract{
Percent-level calibration of cryogenic macro-calorimeters with energy thresholds below 100~eV are crucial for light Dark Matter (DM) searches and reactor neutrino studies based on coherent elastic neutrino-nucleus scattering (CEvNS). This paper presents a novel calibration source based on X-ray fluorescence (XRF) of light elements. It uses a $^{55}$Fe source to irradiate a two-staged target arrangement, emitting characteristic emission lines from 677\,eV to 6.5\,keV. We demonstrate the potential of this new XRF source to calibrate a 0.75 gram CaWO$_4$ crystal of the NUCLEUS and CRAB experiments. Additionally, we introduce CryoLab, an advanced analysis tool for cryogenic detector data, featuring robust methods for data processing, calibration, and high-level analysis, implemented in MATLAB and HDF5. We also present a phenomenological model for energy resolution, which incorporates statistical contributions, systematic effects, and baseline noise, enabling a novel approach to evaluating athermal phonon collection efficiency in macro-calorimeters based on transition edge sensors (TES).}


\maketitle


\footnotetext[5]{Also at LIBPhys-UC, Departamento de Fisica, Universidade de Coimbra, P3004 516 Coimbra, Portugal}
\footnotetext[2]{Now at Max-Planck-Institut für Kernphysik, D-69117 Heidelberg, Germany.}
\footnotetext[3]{Now at Max-Planck-Institut f\"ur Physik, D-85748 Garching, Germany.}
\footnotetext[4]{Now at Dipartimento di Fisica, Universit\`{a} di Milano Bicocca, I-20126, Milan, Italy}

\renewcommand{\thefootnote}{\arabic{footnote}}

\input{Intro}

\input{CalibrationSource}

\input{SetupDataSet}

\input{DataProcessingAnalysis}

\input{XRFanalysis}

\input{ComparisonHeaterXRF}

\input{EnergyResolutionModel}

\input{Conclusion}

\paragraph{Acknowledgments} 
This work has been financed by the CEA, the INFN, the ÖAW and partially supported by the TU Munich and the MPI für Physik. 
NUCLEUS members acknowledge additional funding by the DFG through the SFB1258 and the Excellence Cluster ORIGINS,
by the European Commission through the ERC-StG2018-804228 “NU-CLEUS”, by the P2IO LabEx (ANR-10-LABX-0038) in the
framework "Investissements d’Avenir" (ANR-11-IDEX-0003-01) managed by the Agence Nationale de la Recherche (ANR), France,
by the Austrian Science Fund (FWF) through the "P 34778-N, ELOISE‘, and by the Max-Planck-Institut  für Kernphysik (MPIK), Germany .

\bibliographystyle{sn-mathphys}
\input{xrf.bbl}

\clearpage 
\input{Appendix.tex}

\end{document}

%% file: NucleusAuthorlist.tex
\usepackage{footmisc} 
\renewcommand{\thefootnote}{\fnsymbol{footnote}}

\newcommand{\alsoatLIBPhys}{\P}
\newcommand{\nowatMPIK}{\dag}
\newcommand{\nowatMPP}{\ddag}
\newcommand{\nowatBiccoca}{$\mathsection$}

\author[2]{\fnm{H.} \sur{Abele}~\orcidlink{0000-0002-6832-9051}}
\author[3]{\fnm{G.} \sur{Angloher}}
\author[4]{\fnm{B.} \sur{Arnold}}
\author[6]{\fnm{M.} \sur{Atzori Corona}}
\author[3]{\fnm{A.} \sur{Bento}~\orcidlink{0000-0002-3817-6015}\textsuperscript{\alsoatLIBPhys}}
\author[9]{\fnm{E.} \sur{Bossio}~\orcidlink{0000-0001-9304-1829}}
\author[4]{\fnm{J.} \sur{Burkhart}~\orcidlink{0000-0002-1989-7845}}
\author[5]{\fnm{F.} \sur{Cappella}~\orcidlink{0000-0003-0900-6794}}
\author[5, 8]{\fnm{M.} \sur{Cappelli}~\orcidlink{0009-0002-6148-5964}}
\author[5]{\fnm{N.} \sur{Casali}~\orcidlink{0000-0003-3669-8247}}
\author[6]{\fnm{R.} \sur{Cerulli}~\orcidlink{0000-0003-2051-3471}}
\author[5]{\fnm{A.} \sur{Cruciani}~\orcidlink{0000-0003-2247-8067}}
\author[5]{\fnm{G.} \sur{Del Castello}~\orcidlink{0000-0001-7182-358X}}
\author[5, 8]{\fnm{M.} \sur{del Gallo Roccagiovine}}
\author[2]{\fnm{S.} \sur{Dorer}~\orcidlink{0009-0001-1670-5780}}
\author[1]{\fnm{A.} \sur{Erhart}~\orcidlink{0000-0002-8721-177X}}
\author[4]{\fnm{M.} \sur{Friedl}~\orcidlink{0000-0002-7420-2559}}
\author[4]{\fnm{S.} \sur{Fichtinger}}
\author[4]{\fnm{V. M.} \sur{Ghete}~\orcidlink{0000-0002-9595-6560}}
\author[6, 7]{\fnm{M.} \sur{Giammei}~\orcidlink{0009-0006-9104-2055}}
\author[9]{\fnm{C.} \sur{Goupy}~\orcidlink{0000-0003-4954-5311}\textsuperscript{\nowatMPIK}}
\author[1, 3]{\fnm{D.} \sur{Hauff}}
\author[9]{\fnm{F.} \sur{Jeanneau}~\orcidlink{0000-0002-6360-6136}}
\author[2]{\fnm{E.} \sur{Jericha}~\orcidlink{0000-0002-8663-0526}}
\author[1]{\fnm{M.} \sur{Kaznacheeva}~\orcidlink{0000-0002-2712-1326}}
\author[1]{\fnm{A.} \sur{Kinast}~\orcidlink{0000-0002-5894-2303}}
\author[4]{\fnm{H.} \sur{Kluck}~\orcidlink{0000-0003-3061-3732}}
\author[3]{\fnm{A.} \sur{Langenkämper}}
\author[1, 9]{\fnm{T.} \sur{Lasserre}~\orcidlink{0000-0002-4975-2321}}
\author[9]{\fnm{D.} \sur{Lhuillier}~\orcidlink{0000-0003-2324-0149}}
\author[3]{\fnm{M.} \sur{Mancuso}~\orcidlink{0000-0001-9805-475X}}
\author[2, 9]{\fnm{R.} \sur{Martin}}
\author[3]{\fnm{B.} \sur{Mauri}}
\author[11]{\fnm{A.} \sur{Mazzolari}}
\author[9]{\fnm{L.} \sur{McCallin}}
\author[1]{\fnm{H.} \sur{Neyrial}}
\author[9]{\fnm{C.} \sur{Nones}}
\author[1]{\fnm{L.} \sur{Oberauer}}
\author[1]{\fnm{T.} \sur{Ortmann}\textsuperscript{\nowatMPP}}
\author[10]{\fnm{L.} \sur{Pattavina}~\orcidlink{0000-0003-4192-849X}\textsuperscript{\nowatBiccoca}}
\author*[1, 9]{\fnm{L.} \sur{Peters}~\orcidlink{0000-0002-1649-8582}\textsuperscript{\nowatMPIK}} \email{lilly.peters@tum.de}
\author[3]{\fnm{F.} \sur{Petricca}~\orcidlink{0000-0002-6355-2545}}
\author[1]{\fnm{W.} \sur{Potzel}}
\author[3]{\fnm{F.} \sur{Pröbst}}
\author[3]{\fnm{F.} \sur{Pucci}}
\author[2, 4]{\fnm{F.} \sur{Reindl}~\orcidlink{0000-0003-0151-2174}}
\author[11]{\fnm{M.} \sur{Romagnoni}}
\author*[1]{\fnm{J.} \sur{Rothe}~\orcidlink{0000-0001-5748-7428}} \email{johannes.rothe@tum.de}
\author[1]{\fnm{N.} \sur{Schermer}~\orcidlink{0009-0004-4213-5154}}
\author[2, 4]{\fnm{J.} \sur{Schieck}~\orcidlink{0000-0002-1058-8093}}
\author[1]{\fnm{S.} \sur{Schönert}~\orcidlink{0000-0001-5276-2881}}
\author[2, 4]{\fnm{C.} \sur{Schwertner}}
\author[9]{\fnm{L.} \sur{Scola}}
\author[9]{\fnm{G.} \sur{Soum-Sidikov}~\orcidlink{0000-0003-1900-1794}}
\author[3]{\fnm{L.} \sur{Stodolsky}}
\author[1]{\fnm{R.} \sur{Strauss} ~\orcidlink{0000-0002-5589-9952}}
\author[11, 12]{\fnm{M.} \sur{Tamisari}}
\author[4]{\fnm{R.} \sur{Thalmeier}~\orcidlink{0009-0003-4480-0990}}
\author[5]{\fnm{C.} \sur{Tomei}}
\author[5, 8]{\fnm{M.} \sur{Vignati}~\orcidlink{0000-0002-8945-1128}}
\author[9]{\fnm{M.} \sur{Vivier}~\orcidlink{0000-0003-2199-0958}}
\author[1]{\fnm{A.} \sur{Wex}~\orcidlink{0009-0003-5371-2466}}

\author[1]{\fnm{K.} \sur{v. Mirbach}}
\author[1]{\fnm{V.} \sur{Wagner}~\orcidlink{0000-0003-1845-4951}}

\affil[1]{Physik-Department, Technische Universit\"at M\"unchen, D-85748 Garching, Germany }

\affil[2]{Atominstitut, Technische Universit\"at Wien, A-1020 Wien, Austria}
    
\affil[3]{Max-Planck-Institut f\"ur Physik, D-80805 M\"unchen, Germany}

\affil[4]{Institut f\"ur Hochenergiephysik der \"Osterreichischen Akademie der Wissenschaften, A-1050 Wien, Austria}
   
\affil[5]{Istituto Nazionale di Fisica Nucleare -- Sezione di Roma, Roma I-00185, Italy}
 
\affil[6]{Istituto Nazionale di Fisica Nucleare -- Sezione di Roma "Tor Vergata", Roma I-00133, Italy}
    
\affil[7]{Dipartimento di Fisica, Universit\`{a} di Roma "Tor Vergata", Roma I-00133, Italy}

\affil[8]{Dipartimento di Fisica, Sapienza Universit\`{a} di Roma, Roma I-00185, Italy}
    
\affil[9]{IRFU, CEA, Universit\'{e} Paris-Saclay, F-91191 Gif-sur-Yvette, France}

\affil[10]{Istituto Nazionale di Fisica Nucleare -- Laboratori Nazionali del Gran Sasso, Assergi (L’Aquila) I-67100, Italy}
 
\affil[11]{Istituto Nazionale di Fisica Nucleare -- Sezione di Ferrara, I-44122 Ferrara, Italy}

\affil[12]{Dipartimento di Fisica, Universit\`{a} di Ferrara, I-44122 Ferrara, Italy}

%% file: Intro.tex
\section{Introduction}
\label{intro}

Cryogenic macro-calorimeters achieve ultra-low energy thresholds on the eV-scale \cite{Strauss:2017cam, Ren_2021, Ricochet:2023nvt, Wen_2022}. 
These detectors are extensively used in searches for light Dark Matter (DM)~\cite{CRESST:2019jnq, chang2025, Alkhatib_2021} and in forthcoming neutrino studies based on coherent elastic neutrino-nucleus scattering (CEvNS)~\cite{Angloher:2019flc, Ricochet:2021rjo, Chaudhuri:2022pqk}. An accurate understanding of the detector response is crucial for searching for new physics, such as WIMP interactions or neutrino couplings beyond the Standard Model of particle physics \cite{Billard_2018}.

In this article, we focus on Transition Edge Sensors (TES), which exploit the sharp rise in resistance between the superconducting and normal conducting phases of a thin superconducting film, here tungsten~\cite{Strauss:2017cam, Strauss:2017cuu}. These highly sensitive sensors are used to achieve a low threshold on nuclear recoils by detecting the small heat signals generated by particle interactions within a small crystal. 
While the sensitivity is remarkable, the linear range of TESs is limited and is subject to variations depending on operating conditions such as temperature, magnetic fields, and injected bias current \cite{Irwin2005}.
The measurement results discussed in this article are based on a gram-scale CaWO$_4$ detector of the NUCLEUS collaboration, which aims for an observation of CEvNS from reactor antineutrinos at the Chooz nuclear power plant in France \cite{Angloher:2019flc}.

Typical calibration procedures for cryogenic macro-calorimeters employ $^{55}\mathrm{Fe}$, using its calibration lines at 5.9\,keV and 6.4\,keV, followed by a linear extrapolation of the detector response down to the threshold, spanning two orders of magnitude in energy \cite{PhysRevLett.130.211802}. Such an extrapolation can, in principle, be refined using artificially injected heater pulses of varying and known amplitudes~\cite{Lang_2010}. However, this approach relies on the assumption that particle-induced events and heater pulses produce comparable phonon spectra in both temporal and energy distributions, an assumption that remains difficult to verify experimentally.

Here, we present the development of a new calibration source based on X-ray fluorescence (XRF) of light elements, which addresses these limitations. Our approach employs an initial $^{55}\mathrm{Fe}$ source to irradiate a two-staged target arrangement consisting of readily available XRF targets, including aluminum (Al), glass (\(\mathrm{SiO}_2\)), polytetrafluoroethylene (PTFE), and copper (\(\mathrm{Cu}\)). The resulting XRF source provides a well-defined spectrum of calibration lines ranging from 0.677\,keV (fluorine K$_\alpha$ line) to 6.5\,keV (manganese K$_\beta$ line), enabling detailed characterization across a broad energy range.

This newly developed XRF-based calibration source enables a precise investigation of detector response over a single order of magnitude in energy, eliminating the need for extensive extrapolation over two orders of magnitude down to the detector threshold. Consequently, this approach enhances calibration accuracy and improves the reliability of detector characterization below 1\,keV.
The XRF source produces single-site energy depositions near the crystal surface. It is therefore complementary to the in-situ calibration method of NUCLEUS based on photons emitted by a pulsed LED \cite{Cardani:2021iff}, which provides distributed energy depositions in the bulk.

The remainder of this paper is organized as follows: Section \ref{sec:XRFSource} provides a technical description of the X-ray fluorescence calibration source. In Section \ref{sec:dataset}, we describe the 18.7 days of data taking with the XRF source and the NUCLEUS CaWO$_4$ detector prototype equipped with TES. Section \ref{sec:CryoMeas} covers data taking, processing, and low-level analysis introducing the analysis framework CryoLab. In Section \ref{sec:xrfanalysis}, we present the high-level analysis of the XRF data, including the derivation of the calibration curve and spectrum extraction. Section \ref{sec:compheaterxrf} presents the comparison of XRF data with artificial heater pulses. Section \ref{sec:eneryresolutionmodel} discusses applications of the XRF source in modeling the detector energy resolution. Finally, we conclude with an outlook in Section \ref{sec:Outlook}, highlighting the interplay of different calibration methods.

%% file: CalibrationSource.tex
\section{X-ray Fluorescence Calibration Source} 
\label{sec:XRFSource} 

Low-energy radioactive sources are commonly used for the calibration of cryogenic detectors at energies of few keV. A prominent example is the radioisotope $^{55}$Fe, which provides low-energy X-rays. The isotope decays by electron capture and emits the characteristic fluorescence X-rays of the daughter nucleus $^{55}$Mn at $\SI{5.9}{keV}$ (K$_\alpha$) and  $\SI{6.5}{keV}$ (K$_\beta$).  X-ray fluorescence can be used to produce additional calibration lines at lower energies, useful to calibrate detectors sensitive to smaller energies and to study the linearity of the detector response. To achieve this, a radioactive source is used to excite inner shell electrons of a low-Z material which subsequently emits its characteristic X-rays. This technique has been successfully used for detector calibration with $^{55}$Fe primary sources illuminating aluminum targets ($E_{K_\alpha}=\SI{1.49}{keV}$) facing the detector, either in reflection~\cite{COLLING1995408} or in transmission (using thin foils and membranes)~\cite{PhysRevLett.127.061801, LEBLANC1996208}. 

\begin{table}[]
    \centering
     \caption{Summary of the X-ray energies and fluorescence yield y$_\text{fluor}$ for elements Z$\leq$10, and selected elements relevant for our XRF calibration source~\cite{XrayDB}. For the line energy, we give the weighted average over all K-shell transitions. For elements emitting several relevant lines, we give their intensity in relative units.}
    \begin{tabular}{l|c|c|c|c}
        Element & Line & Energy [eV] & y$_\text{fluor}$ [\%] & rel. Intensity \\
         \hline
         C  & K$_{\alpha}$ &  277.0 &  0.140 &  \\ \hline
         O  & K$_{\alpha}$ &  524.9 &  0.580 &  \\ \hline
         F  & K$_{\alpha}$ &  676.8 &  0.920 &  \\ \hline
         Cu & L$_{\alpha}$ &  927.7 &  1.100 & \\ \hline 
         Al & K$_{\alpha}$ & 1486.4 &  3.298 & 0.888\\
         Al & K$_{\beta}$  & 1557.0 &  3.298 & 0.112\\ \hline
         Si & K$_{\alpha}$ & 1739.6 &  4.291 & 0.887 \\
         Si & K$_{\beta}$  & 1837.0 &  4.291 & 0.113 \\ \hline
         K  & K$_{\alpha}$ & 3312.9 & 13.207 & 0.882 \\
         K  & K$_{\beta}$  & 3590.1 & 13.207 & 0.118 \\ \hline
         Ca & K$_{\alpha}$ & 3691.1 & 14.679 & 0.881\\
         Ca & K$_{\beta}$  & 4013.1 & 14.679 & 0.119 \\ \hline
         Mn & K$_{\alpha}$ & 5896.5 & 31.918 & 0.876 \\
         Mn & K$_{\beta}$  & 6492.1 & 31.918 & 0.124 \\  \hline
         Cu & K$_{\alpha}$ & 8039.6 & 44.109 & 0.872 \\ 
         Cu & K$_{\beta}$  & 8903.6 & 44.109 & 0.128 \\        
    \end{tabular}
    \label{tab:xrf_lines_yields}
\end{table}

Extending this calibration technique to lower X-ray energies, emitted by lower-Z materials, is challenging for several reasons. The photo-ionisation cross-section and fluorescence yield, defined as the number of emitted fluorescence X-rays per excited shell vacancy, drop steeply for lower $Z$ targets (see Table~\ref{tab:xrf_lines_yields}). This makes it difficult to produce low-energy XRF peaks of sufficient rates for calibration. In addition, for light XRF targets there is increasing competition from the production of Auger electrons, which produce unwanted background in cryogenic detectors, covering the low-rate calibration lines.

In a practical calibration source, the emitted calibration lines should span a large energy range, yet be produced with similar intensities. The total rate of the source is limited by the response time of the cryodetector, to avoid excessive pile-up. The rate of the least intense calibration line then determines the exposure time necessary to achieve sufficient statistics across the source spectrum. Typically, low-energy XRF lines are emitted with a very low rate, requiring prohibitively long exposure times.

\begin{figure}[h!]
    \centering
    \includegraphics[width=0.45\textwidth]{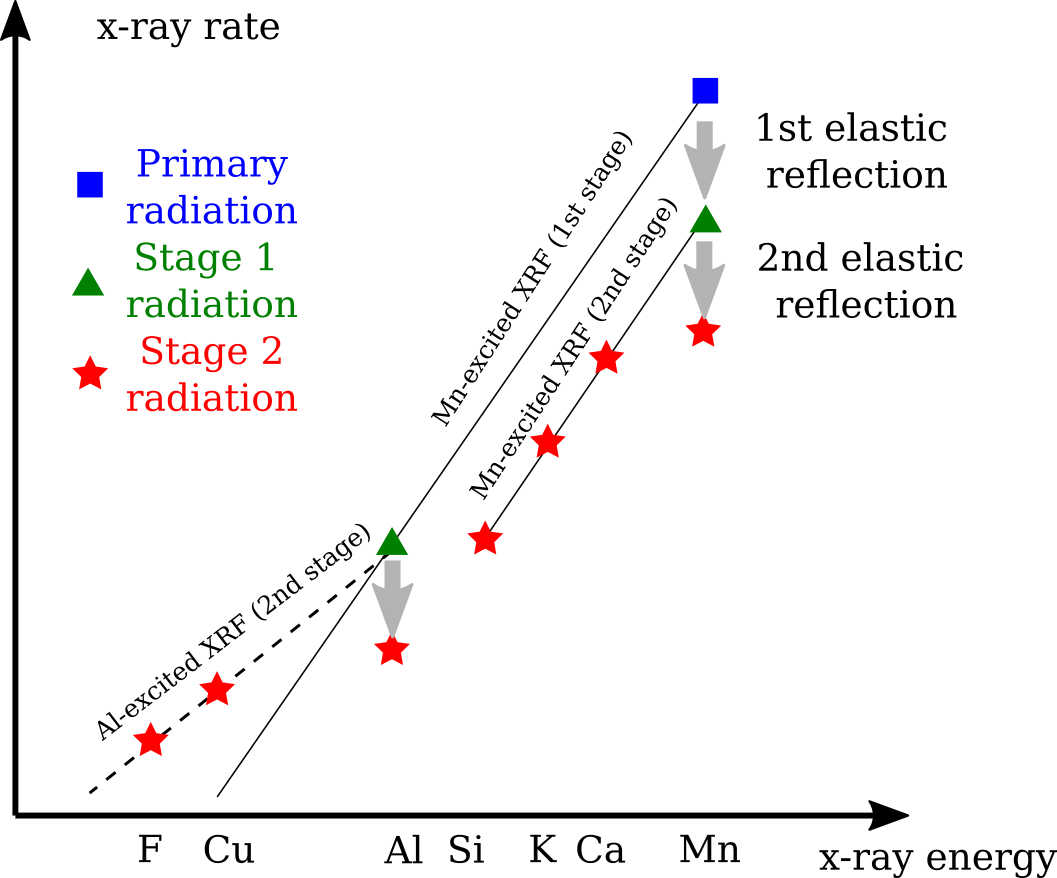}
    \caption{Conceptual scheme of the two-stage X-ray fluorescence source.}
    \label{fig:source_concept}
\end{figure}

In this work, we developed a novel configuration for an XRF source which addresses these challenges. The design uses a two-stage X-ray fluorescence configuration: a first target of intermediate X-ray energy is excited by primary radiation, and faces a second target for production of low-energy fluorescence. This takes advantage of the larger photo-ionisation cross-section for lower energy of the exciting radiation, to increase the fluorescence output of the second stage. 
Both XRF targets are used in reflection at $45^\circ$-angles to the direction of incidence. Thus, primary radiation has to undergo two elastic reflections to reach the detector, and the intermediate-energy fluorescence produced in the first target has to undergo one elastic reflection. This serves to suppress the on-detector rates of higher-energy X-rays relative to the fluorescence of the second target, reducing the exposure time needed to obtain a source spectrum.
As the second target is illuminated both by reflected primary radiation and first-stage fluorescence, a multi-element target allows to additionally excite X-rays above the fluorescence energy of the first stage target.

Fig.~\ref{fig:source_concept} illustrates the processes by which X-ray fluorescence is produced and transported in the two-stage source. It shows the main conceptual features: suppression of higher energy rates by reflections, and enhancement of low-energy fluorescence by reduced excitation energy. The schematic does not quantitatively account for factors such as different solid angles, photoionization cross-sections, and XRF yields.

\begin{figure}[h!]
    \centering
    \includegraphics[width=0.45\textwidth]{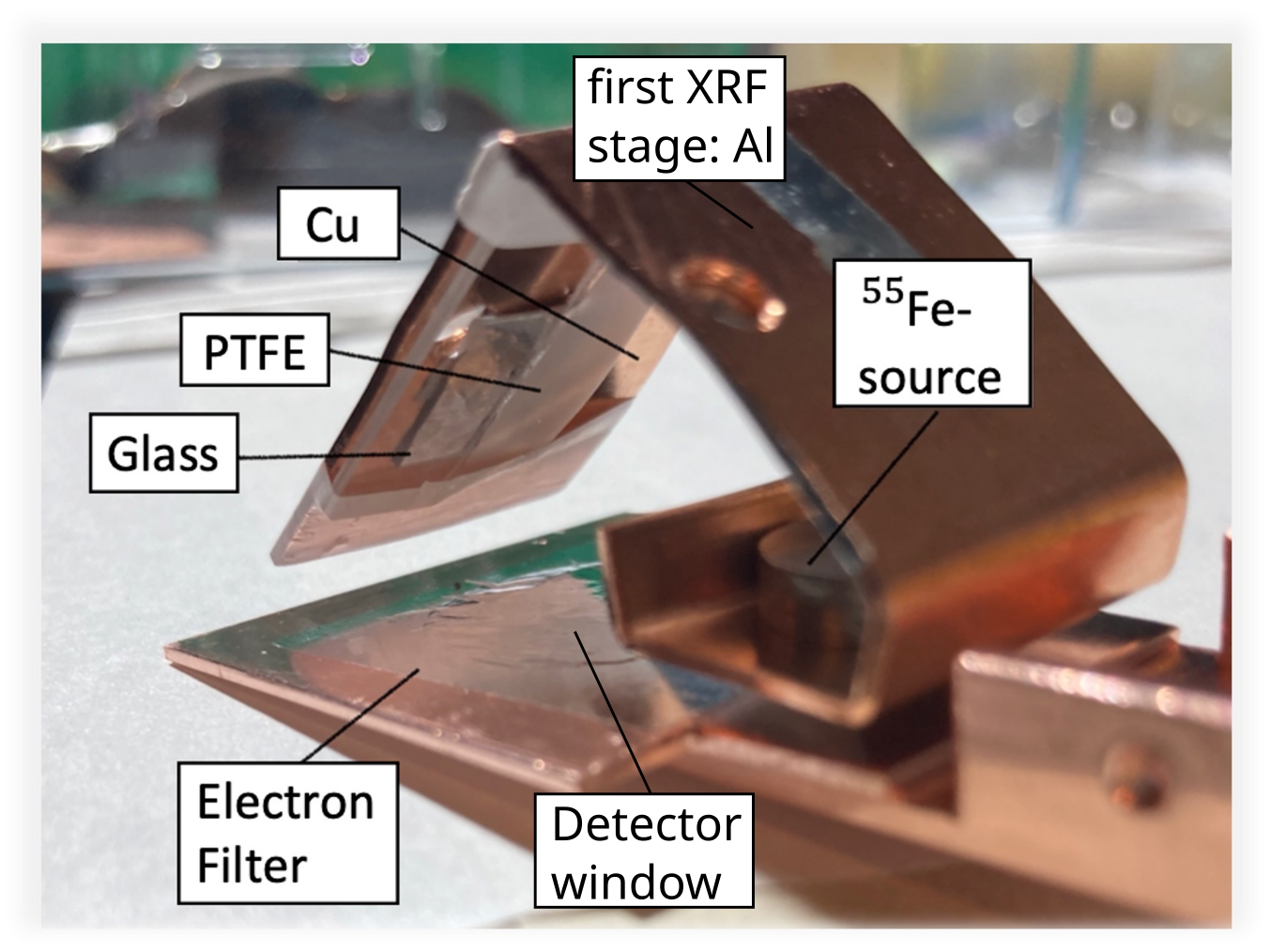}
    \caption{Photograph of the two-stage X-ray fluorescence (XRF) source used for calibrating the CaWO$_4$ detector crystal. The setup features a primary \(^{55}\)Fe radioactive source, which illuminates a first XRF stage formed by aluminum tape. This in turn faces a secondary XRF stage composed of PTFE, glass, and copper (also serving as the supporting structure). An electron filter formed by a $\mu$m-thin aluminum foil covers the detector window to ensure that only the desired X-rays reach the detector. During the measurement, the detector is placed directly below the detector window.}
    \label{fig:source}
\end{figure}

Our implementation of this two-stage source concept is shown in Fig.~\ref{fig:source}. The primary radiation is provided by an $^{55}$Fe source ($\SI{12.5}{mm}$ diameter, $\SI{3.5}{mm}$ thickness, $\SI{2.25e5}{Bq}$ activity at the time of measurement). The first target is formed by a square piece of aluminum tape with a size of $\SI{22}{mm}\times\SI{22}{mm}$. The choice of aluminum as first stage XRF target, with an intermediate fluorescence energy around $\SI{1.5}{keV}$, is a compromise between the first stage fluorescence yield, which drops for lower $Z$ targets, and the photo-ionisation cross-sections for the second stage, which increase for lower $Z$ targets. 

The second target consists of a $ \SI{7}{mm} \times \SI{25}{mm}$ glass slab (for Si, Ca and K fluorescence lines), a $ \SI{7}{mm} \times \SI{25}{mm}$ PTFE piece (F fluorescence), and an exposed patch of the copper structure of similar dimension (for the L$_\alpha$ line). Finally, an aluminum foil (thickness $\sim\SI{1}{\mu m}$) blocks Auger electrons from reaching the cryodetector while transmitting a large fraction of the low-energy XRF photons. It was produced by electron-beam evaporation on top of a photo-resist layer, and transferred to the copper holder of the XRF source in a solvent bath.
The housing made from tempered copper has  a width of $\SI{22}{mm}$, length of $\SI{58}{mm}$ and height of $\SI{28}{mm}$. During operation the structure is closed by lateral walls.

%% file: SetupDataSet.tex
\section{Setup and Dataset}
\label{sec:dataset}

The experimental setup discussed in this article was located in an above-ground laboratory at the Physics Department of the Technical University of Munich. This setup was designed specifically for the calibration of the CaWO$_4$ detector crystal coupled to a TES during the prototyping phases of the NUCLEUS experiment.
It is important to note that this configuration is not the final design chosen for the initial phase of the NUCLEUS experiment at the Chooz EDF nuclear power station.

The detector is housed within a Bluefors LD400 dry dilution refrigerator \cite{Bluefors}, which maintains a base temperature below 7\,mK. This type of cryostat employs a pulse tube cryocooler to provide a temperature bath below $\SI{4}{K}$. This eliminates the need for cryogenic liquid refills and allows maintaining stable temperature conditions at the detector for uninterrupted data acquisition. To minimize mechanical coupling between cryostat and detectors and resultant vibrational noise from the pulse tube, the detector assembly is isolated using a custom vibration decoupling system \cite{VDS2025}.

The detector is a cubic CaWO$_4$ crystal of $\SI{0.75}{g}$ mass, positioned between small sapphire spheres and secured using a flexible bronze clamp within a copper support box. Thermal connectivity between the crystal and the support structure is facilitated by a gold wirebond, ensuring heat dissipation to the copper holder, which serves as a thermal bath.

The CaWO$_4$ detector is equipped with a tungsten (W) thin-film TES \cite{Strauss:2017cuu}, operating at a superconducting transition temperature of 14\,mK. For stable operation, the TES is biased with a constant current of $\SI{3}{\mu A}$, and its operating resistance is regulated by a resistive heater formed by a thin gold film adjacent to the TES structure. 

The XRF radioactive source, detailed in Section \ref{sec:XRFSource}, is positioned above the copper box, emitting X-rays that interact with the CaWO$_4$ crystal to enable energy calibration.

\begin{figure}[h!]
    \centering
    \includegraphics[width=0.45\textwidth]{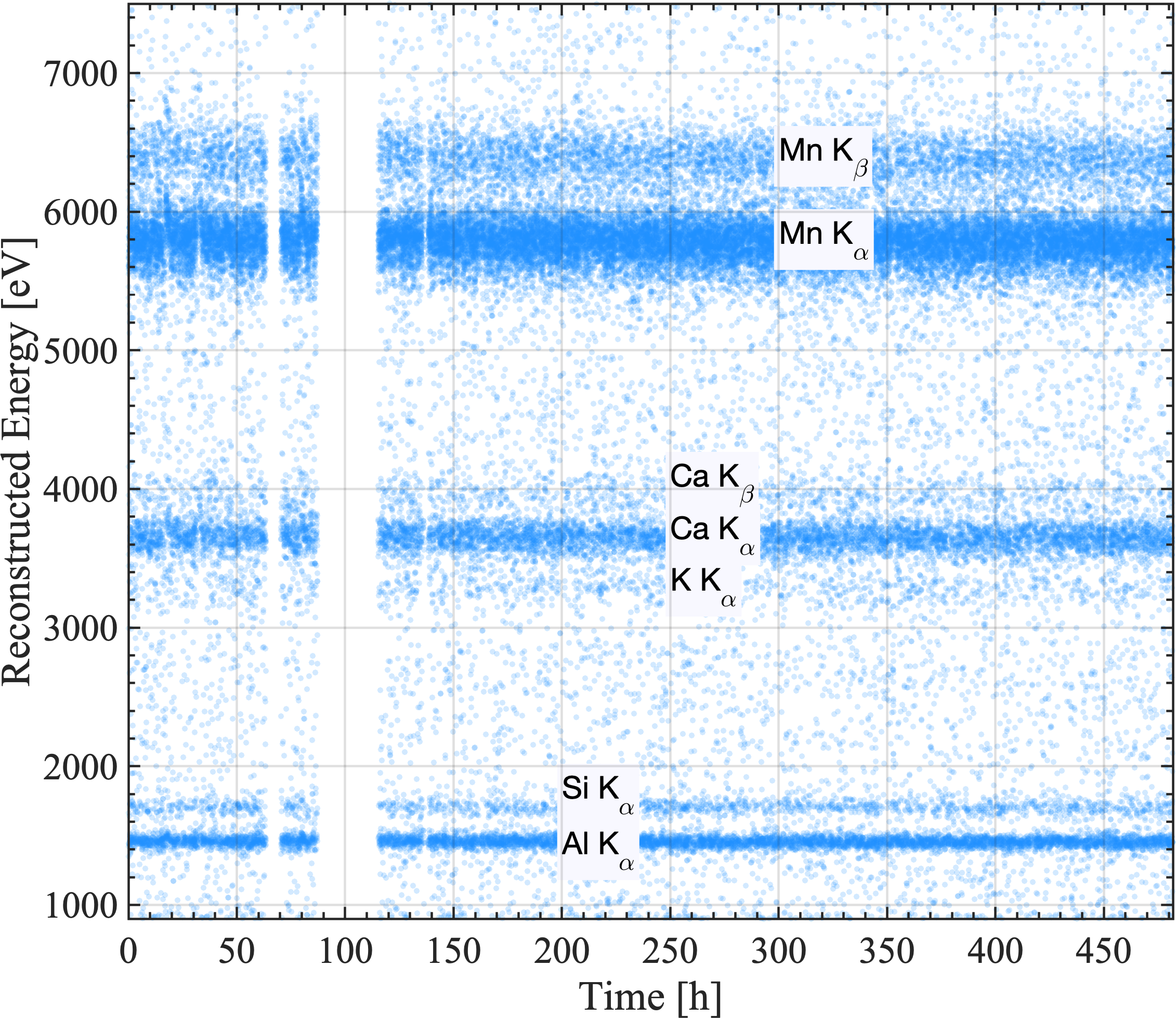}
    \caption{Reconstructed energy (eV) versus time (hours) for 450 hours of measurement using the optimal filter technique. Each point represents a detected pulse. The X-ray lines for Mn K$_\alpha$, Mn K$_\beta$, Ca K$_\alpha$, Ca K$_\beta$, K K$_\alpha$, Si K$_\alpha$, and Al K$_\alpha$ are highlighted. The detector stability is on the sub-percent level over the entire measurement period. Details of the analysis and calibration will follow in the next sections.}
    \label{fig:stability}
\end{figure}

TES signals are read out using a commercial low-$T_c$ DC-SQUID system (STAR Cryoelectronics ``S"-Series SQ100)~\cite{starcryo}. The data stream was acquired using the custom VDAQ02 data acquisition system developed for the cryogenic experiments CRESST~\cite{CRESST:2019jnq}, NUCLEUS~\cite{Angloher:2019flc} and COSINUS~\cite{PhysRevD.110.043010}, operating at a sampling frequency of $\SI{10}{kHz}$. Data were recorded on disk as binary files, with each 100-microsecond sample stored as an int16 binary word. The measurement campaign ran from February 17th to March 10th, 2023, accumulating 18.67 days (448 hours) of continuous data acquisition. There were minor interruptions totaling approximately 10\% of the overall time, as displayed in Fig. \ref{fig:stability}.

%% file: DataProcessingAnalysis.tex
\section{Data Processing and Low Level Analysis}
\label{sec:CryoMeas}

\begin{figure*}[h!]
    \centering
    \includegraphics[width=1.\textwidth]{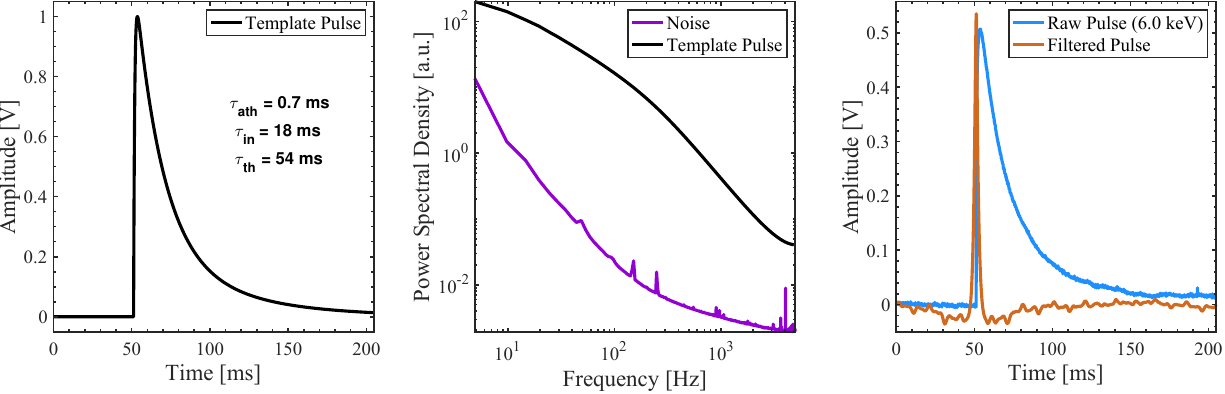}
    \caption{Illustration of the optimal filter amplitude reconstruction. Left: Template pulse (black) without noise in the time domain, the rise and decay times according to \cite{Probst:1995hjq} are indicated. Center: Template pulse (black) and noise (purple) power spectral density in the frequency domain, where the ratio defines the transmission function of the filter. Right: A 6.1 keV raw pulse with noise in the time domain (blue) and the filtered time series (orange).}
    \label{fig:ofreco}
\end{figure*}

Data analysis was performed using the CryoLab tool, developed collaboratively by TU Munich and CEA Saclay. This MATLAB-based tool~\cite{matlab} offers a modular, scalable framework for processing complex datasets from cryogenic detectors such as small CaWO$_4$ and Al$_2$O$_3$ crystals, operated at ultra-low temperatures (20 mK), equipped with TES, and read out via SQUID electronics. 

CryoLab handles continuous data streams from multiple detectors or hardware-triggered data in various formats produced by customized DAQ systems in use by the NUCLEUS collaboration.
In parallel to actual data analysis, stream simulations generate synthetic data for testing and validating each step of the analysis pipeline.

Once the stream is processed and triggering performed, all timestamps, pulses, noise traces, and relevant analysis information are stored in HDF5 files \cite{hdf5}, a hierarchical data format supporting efficient storage and retrieval of large, complex datasets. This centralization facilitates organized data management and subsequent analysis steps.
At each step of the analysis process, the HDF5 files are updated with newly calculated quantities, maintaining a continuous and comprehensive record.

Detected pulses are modeled to extract parameters such as amplitude, rise time and two decay times for both thermal and athermal phonon signal components~\cite{Probst:1995hjq}. Pulse shape fitting is used to determine the pulse shape describing the data best (template pulse), followed by the application of the optimal filter. The optimal filter technique~\cite{Gatti:1986cw} enhances faint signal components and suppresses noise, thereby maximizing the signal-to-noise ratio and improving energy resolution. It uses the template pulse shape obtained by fitting the model~\cite{Probst:1995hjq} to each pulse and averaging the fit parameters,  and the averaged noise power spectrum from randomly triggered stream samples acquired during data collection and cleaned from accidental pulses.

CryoLab applies straightforward  data cleaning procedures to remove irrelevant triggers, conducted in two levels. Level 1 involves assessing the stream slope for pile-up rejection, eliminating spurious events such as flux quantum losses in SQUIDs, evaluating the standard deviation and absolute baseline, and identifying traces with multiple peaks. Level 2 utilizes pulse shape fit parameters and chi-square, or the mean square error of the optimal filter reconstruction. Both levels of cuts are tuned and evaluated using stream simulations closely mimicking the actual data.

Additionally, anticoincidence cuts with external vetos may be applied to remove background events. Efficiency corrections may be made to account for detector performances.

Calibration is a crucial component of the CryoLab analysis chain and typically \(^{55}\text{Fe}\) sources are used for preliminary calibration, establishing a relationship between energy and signal voltage at 5.9 keV.

During data acquisition, CryoLab performs an automated rapid preliminary analysis after each 12-hour measurement interval, enabling monitoring and validation of the ongoing data collection. The analysis results are semi-automatically processed and updated on a web server.
This approach ensures that any anomalies are promptly detected and addressed, maintaining the integrity and reliability of the dataset.

Figure \ref{fig:ofreco} illustrates the process of optimal filter amplitude reconstruction for the XRF measurement presented here. The left panel displays a template pulse\footnote{The template pulse is extracted from the energy region of the Al K$_{\alpha}$ line. No energy dependence of the pulse shape is observed.} corresponding to the pulse shape of the expected signal without noise. The middle panel shows the signal and noise power spectrum in the frequency domain, where the ratio defines the transmission function of the filter. The right panel presents an example of the raw pulse with noise in the time domain, and the corresponding filtered time series.
After filtering, a baseline resolution\footnote{The baseline resolution is estimated by applying the optimal filter to a set of noise traces. The noise amplitude at a fixed position follows a Gaussian distribution, and its standard deviation represents the baseline resolution estimate.} of $\sigma_0 = \SI{1.03(4)}{mV}$ is achieved.
The calibration method for this XRF measurement taking advantage of the multiple X-ray lines will be addressed in the next section. 
The quadratic calibration as described in the next section leads to a baseline resolution in energy units of $\sigma_0 = \SI{9.9(4)}{eV}$ in this measurement.

The efficiency of the optimal filter-based offline triggering algorithm was studied and found to be approximately 90\% and nearly constant in the energy region of interest for the XRF analysis.

%% file: XRFanalysis.tex
\section{Sub-keV Electron Recoil Calibration of Macroscopic Cryogenic Calorimeters Using X-rays}
\label{sec:xrfanalysis}

The primary objective of this analysis is to achieve sub-keV electron recoil (ER) calibration for cryogenic detectors taking advantage of the multiple X-ray lines produced by the XRF source
described above.
The procedure aims to relate the detector's optimal filter amplitude (in mV) to the true energy over a broad range, ideally from a few tens of eV to approximately 10 keV.

We begin by identifying the most prominent X-ray lines in the energy spectrum: \(\mathrm{F}\) K$\alpha$ (0.677\,keV), \(\mathrm{Cu}\) L$\alpha$ (0.927\,keV), \(\mathrm{Al}\) K$\alpha$ (1.486\,keV), \(\mathrm{Si}\) K$\alpha$ (1.740\,keV), \(\mathrm{K}\) K$\alpha$ (3.313\,keV), \(\mathrm{Ca}\) K$\alpha$ (3.691\,keV), \(\mathrm{Ca}\) K$\beta$ (4.013\,keV), \(\mathrm{Mn}\) K$\alpha$ (5.897\,keV), \(\mathrm{Mn}\) K$\beta$ (6.492\,keV), and \(\mathrm{Cu}\) K$\alpha$ (8.040\,keV). The \(\mathrm{Cu}\) K$\alpha$ line originates from muon-induced X-ray fluorescence.

Local fits of the uncalibrated spectrum around each identified peak are performed using single, double, or triple Gaussian models, according to the number of peaks in the selected fit window (see Fig. \ref{fig:peakfit_FCu}-\ref{fig:peakfit_Cu8keV} in the appendix). Each fit provides the peak location (in mV), the standard deviation (resolution), and a constant background. The optimal filter reconstruction locations and resolutions (in mV) are then correlated with the true energies (in eV) obtained from X-ray databases. Results for each X-ray line used in further analysis are provided in Table \ref{tab:xray_results}.

\begin{table*}[h!]
    \centering
    \begin{tabular}{c|c|c|c|c|c}
    X-ray Line & Energy [eV] & Rate [cpd] & OF Amplitude [mV] & eV/V & Resolution [mV] \\ \hline 
    F K$_\alpha$ & 676.8 & 6.03 $\pm$ 0.65 & 69.1 $\pm$ 0.1 & 9797.9 $\pm$ 17.5 & 2.1 $\pm$ 0.3 \\ \hline 
    Cu L$_\alpha$ & 927.7 & 15.96 $\pm$ 1.05 & 94.7 $\pm$ 0.2 & 9797.9 $\pm$ 17.5 & 2.1 $\pm$ 0.3 \\ \hline 
    Al K$_\alpha$ & 1486.4 & 315.68 $\pm$ 4.69 & 150.2 $\pm$ 0.0 & 9895.4 $\pm$ 2.7 & 2.9 $\pm$ 0.3 \\ \hline 
    Si K$_\alpha$ & 1739.6 & 62.47 $\pm$ 2.09 & 174.7 $\pm$ 0.1 & 9957.8 $\pm$ 7.6 & 3.1 $\pm$ 0.3 \\ \hline 
    K K$_\alpha$ & 3312.9 & 45.88 $\pm$ 1.79 & 323.2 $\pm$ 0.4 & 10248.9 $\pm$ 11.7 & 6.9 $\pm$ 0.8 \\ \hline 
    Ca K$_\alpha$ & 3691.1 & 371.29 $\pm$ 5.09 & 354.4 $\pm$ 0.1 & 10416.3 $\pm$ 2.8 & 6.7 $\pm$ 0.7 \\ \hline 
    Ca K$_\beta$ & 4013.1 & 61.68 $\pm$ 2.07 & 381.4 $\pm$ 0.4 & 10521.3 $\pm$ 9.7 & 7.1 $\pm$ 0.8 \\ \hline 
    Mn K$_\alpha$ & 5896.5 & 1317.78 $\pm$ 9.58 & 528.0 $\pm$ 0.1 & 11167.0 $\pm$ 1.5 & 10.1 $\pm$ 1.0 \\ \hline 
    Mn K$_\beta$ & 6491.8 & 252.13 $\pm$ 4.19 & 571.4 $\pm$ 0.2 & 11361.3 $\pm$ 4.0 & 9.0 $\pm$ 0.9 \\ \hline 
    Cu K$_\alpha$ & 8039.7 & 23.30 $\pm$ 1.27 & 681.7 $\pm$ 0.8 & 11794.1 $\pm$ 13.6 & 10.6 $\pm$ 1.4 \\
    \end{tabular}
    \caption{Summary of the prominent X-ray lines in the XRF spectrum, including their literature energies \cite{XrayDB}, the measured rates, reconstructed amplitudes, calibration values, and resolutions. The results are obtained through local fits using simple Gaussian models.}
    \label{tab:xray_results}
\end{table*}

\begin{figure}[h!]
    \centering
    \includegraphics[width=0.45\textwidth]{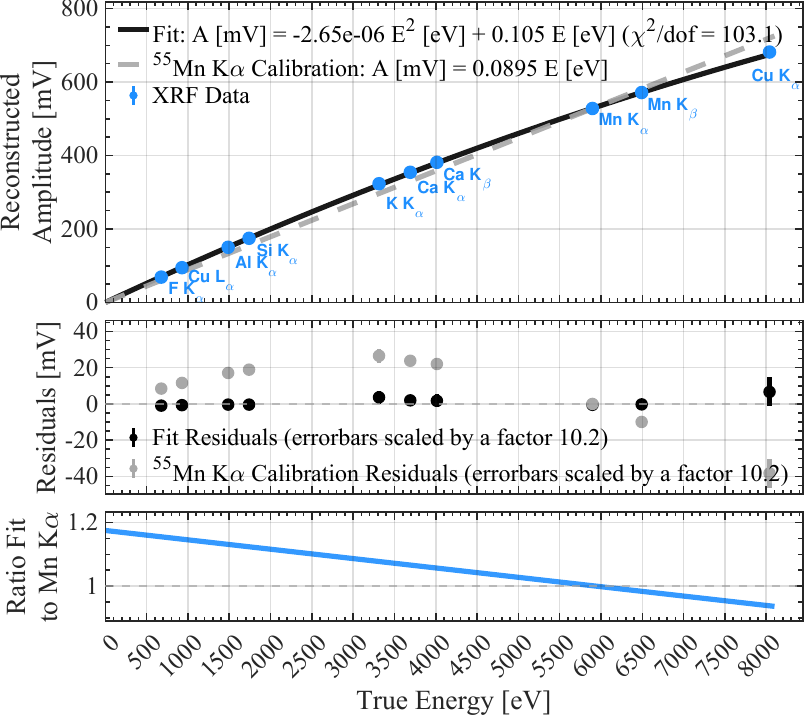}
    \caption{Top panel: Reconstructed amplitude in mV of identified X-ray lines from XRF measurement as a function of true energy in eV. The blue points represent the measured data, the error bars are too small to be displayed. Two calibration curves are shown: (a) a linear interpolation anchored at the Mn K$_\alpha$ line and at zero (grey dashed), and (b) a quadratic fit using all data points (black solid). The quadratic model is strongly favored, with the best-fit formula given in the figure: \(A \, [\text{mV}] = -2.65 \times 10^{-6} E^2 \, [\text{eV}] + 0.105 E \, [\text{eV}] \) with \(\chi^2/\text{dof} =824.8/8\).
    Middle panel: Residuals of the data minus the model fits (linear in gray and quadratic in black) as a function of true energy. Here we normalize the reduced chi-squared (\(\chi^2/\text{dof}\)) to unity.
    Bottom panel: Ratio of the quadratic fit function to the linear calibration, illustrating the error in the energy scale that would result from using the incorrect linear model anchored at the Mn K$_\alpha$ line.} 
    \label{fig:calcurve}
\end{figure}

The results are depicted in Fig. \ref{fig:calcurve}. The reconstructed amplitude in mV of the identified X-ray lines is plotted as a function of true energy in eV. The blue points represent the data. Given the excellent statistics in most of the peaks, the statistical uncertainty on the peak positions is generally very small (on the order of  0.01\%) leading to a relatively high $\chi^2/\text{dof}$ value dominated by systematic deviations from the quadratic fit.
Two calibration curves are shown: (a) a linear extrapolation anchored at the Mn K$_\alpha$ line and zero, and (b) a quadratic fit using all data points. The quadratic model is favored, with the best-fit formula  $A \, [\text{mV}] = -2.65 \times 10^{-6} E^2 \, [\text{eV}] + 0.105 E \, [\text{eV}]$, with \(\chi^2/\text{dof} = 103.1\).
Therefore, the uncertainties are scaled to normalize the \(\chi^2\) per degree of freedom to unity\footnotemark.
\footnotetext{We scale the error bars by incorporating an ad-hoc systematic error. Given \(\chi^2_{\text{target}} = 8\), \(\chi^2_{\text{initial}} = 103.1\), and 8 degrees of freedom, the scaling factor is \(k = \sqrt{824.8/8} \approx 10.2\). The total uncertainty is adjusted as \(\sigma_{\text{total}} = 10.2 \cdot \sigma_{\text{stat}}\). This is equivalent to adding a systematic uncertainty \(\sigma_{\text{sys}} = \sqrt{(10.2 \cdot \sigma_{\text{stat}})^2 - \sigma_{\text{stat}}^2} \approx 10.2 \cdot \sigma_{\text{stat}}\).} Still it is noteworthy that even after this normalization the uncertainties are of the order of only 0.1\%. A possible systematic error could originate from the uncertainty in the true line energies, due to chemical shifts or satellite lines \cite{hell2025}.

The middle panel of Fig. \ref{fig:calcurve} shows the residuals of the data minus the models (linear in gray and quadratic in black) with rescaled errorbars as a function of true energy.

The bottom panel of Fig. \ref{fig:calcurve} presents the ratio of the quadratic fit to the linear model, illustrating the error in the energy scale resulting from using the incorrect linear model. This error reaches a maximum of 18\% at baseline.

Given the fit results, the quadratic model provides a relatively better description of the data. The linear fit introduces significant distortions, whereas the quadratic fit, although affected by systematic uncertainties, offers a more accurate calibration.
We performed a chi-square test to assess the validity of the linear hypothesis, where the energy calibration is modeled as a straight line anchored at zero and the Mn K$_\alpha$ line energy. The quadratic fit to the XRF data was used as an alternative hypothesis. We found that $\Delta \chi^2$ is approximately 2730 (with $\Delta$dof = 1), strongly disfavoring the linear hypothesis. Given this large $\Delta \chi^2$, the corresponding p-value is expected to be exceedingly small. However, at such extreme values, accurately determining the significance level is challenging. Nevertheless, the result clearly excludes the linear hypothesis at a confidence level exceeding 5$\sigma$ by far.

\begin{figure}[h!]
    \centering
    \includegraphics[width=0.45\textwidth]{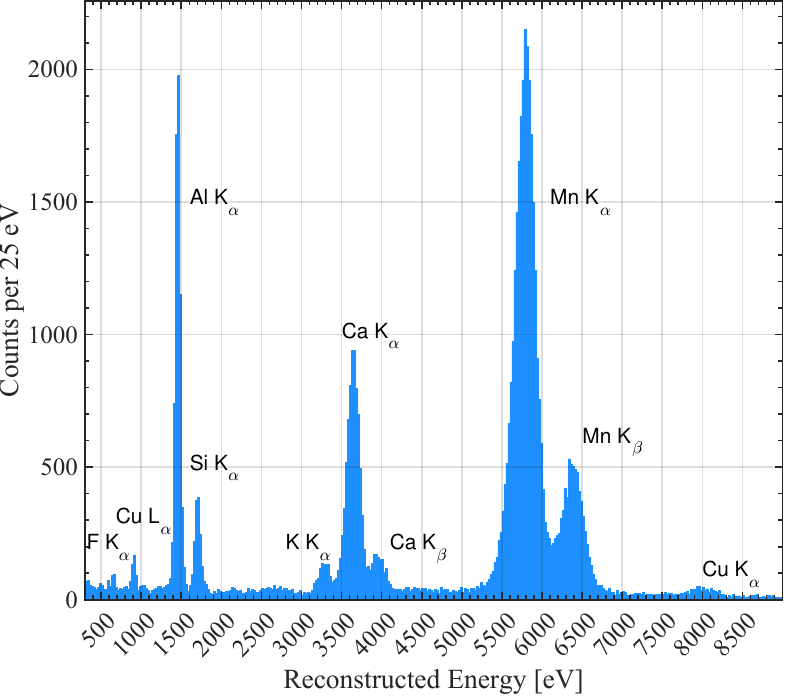}
    \caption{Calibrated energy spectrum obtained from XRF measurements, using the quadratic calibration to map voltage (V) to the actual X-ray energies. 
    Each peak represents a specific elemental transition as listed in the text, validating the precise energy calibration achieved through CryoLab's processing and analysis.}
    \label{fig:spectrum}
\end{figure}

The calibration curve helps to understand the relationship between the detector response (measured in mV) and the true energy of incident X-rays, which is crucial for accurately interpreting the experimental data. We adopt the quadratic calibration and plot the calibrated energy spectrum in Fig. \ref{fig:spectrum}. The spectrum shows prominent peaks corresponding to the X-ray transitions listed above. 

The XRF calibration is particularly valuable because it enables mapping the detector response across a broad energy range, from about 500 eV to approximately 10 keV. However, it is crucial to recognize that comprehensive detector calibration demands a considerable amount of exposure. For example, as shown in Table \ref{tab:xray_results}, accumulating 1000 counts requires roughly 18 hours for the Mn K$_\alpha$ peak, 75 hours for the Al K$_\alpha$ peak, and about 4000 hours for the F K$_\alpha$ peak below 1\,keV. Consequently, while a few days of data acquisition suffice to map the detector response down to 1\,keV, reaching the sub-keV range necessitates several weeks of data collection with the current XRF source.

%% file: ComparisonHeaterXRF.tex
\section{Comparison with Artificial Heater Pulses}
\label{sec:compheaterxrf}

The TES temperature is regulated by an actively controlled resistive heater, consisting of a 50 nm gold film at the surface of the CaWO$_4$ detector crystal. Both DC heater current and fast heater pulses can be injected into the system. The stabilization process involves the periodic injection of heater pulses with varying amplitudes. Saturated heater pulses, known as ``control pulses", provide a direct measurement of the TES resistance within its superconducting transition range. These control pulse amplitudes are used in a PID loop controlling the DC heater current, stabilizing the detector against slow temperature drifts over a time scale of tens of seconds.

Smaller heater pulses, referred to as ``test pulses", ranging from near-threshold levels to detector saturation can be applied. These pulses are always used to monitor the temporal stability of the detector's energy scale and can also additionally serve to linearize the energy response during post-processing. As the power dissipated in the heater is proportional to the square of the applied heater current, heater pulses with a known ratio of injected energies can be constructed. By injecting such test pulses with energies known up to a scaling factor, the reconstructed amplitudes allow for verification and correction of linearity, facilitating a more accurate calibration of the detector, assuming that they behave like particle pulses.

This section compares the detector response to XRF pulses with that of heater pulses to determine whether heater pulses can effectively serve as a proxy for calibrating against actual XRF events, thereby ensuring the accuracy of the energy scale calibration. For the first time, we present a comprehensive map of X-ray induced signals, from sub-keV to approximately 9 keV, alongside a corresponding set of test pulses of similar amplitudes. This enables a detailed comparison and investigation of the test pulse linearization of the energy scale.

\begin{figure}[h!]
    \centering
    \includegraphics[width=0.45\textwidth]{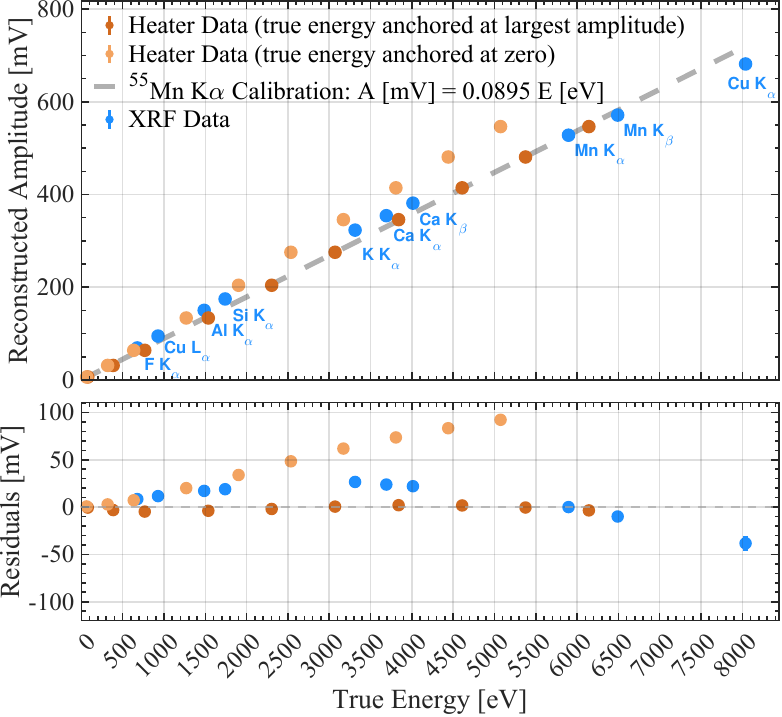}
    \caption{
    Top panel: Reconstructed amplitude in mV of identified X-ray lines from XRF data (blue) compared to heater data (orange) as a function of true energy in eV.
    True energies for the heater data were assigned by assuming a linear relation between injected heater energy and true energy and matching the test pulse amplitude-energy relation to the amplitude-energy relation of the XRF data, either at the Mn K$_\alpha$ line (dark orange) or at low energies (light orange). The grey dashed line represents the linear interpolation anchored at the Mn K$_\alpha$ line and at zero.
    Bottom panel: Residuals of the XRF and heater data relative to the Mn K$_\alpha$ calibration as a function of true energy.
    }
    \label{fig:calcurve-heater}
\end{figure}

Figure \ref{fig:calcurve-heater} illustrates the comparative analysis between XRF pulses and heater pulses. In the top panel, the reconstructed amplitudes (in mV) of X-ray lines from XRF data (blue) are compared to those from heater data (orange) as a function of true energy (in eV). The conversion to true energies for the heater data is ambiguous and was performed based on the largest heater pulse amplitude (dark orange) or by extrapolating the heater pulse amplitudes to zero (light orange) using the quadratic fit from figure 
\ref{fig:calcurve}. The grey dashed line represents the linear interpolation anchored at the Mn K$_\alpha$ line and zero. While there is a global agreement within 20\% across the entire energy range, we observe a distinct difference in behavior: XRF pulses exhibit a clear non-linearity, whereas heater pulses maintain a linear trend with energy.

In previous measurements it has been demonstrated that particle and heater pulses share identical non-linear amplitude energy relation over more than an oder of magnitude in energy~\cite{Lang_2010}, by using an appropriate energy reconstruction algorithm.  
The underlying cause of the discrepancy in this measurement remains unclear, suggesting that caution is necessary when using heater pulses for energy scale linearization.
 Firstly, the pulse height reconstruction method based on the Optimal Filter algorithm~\cite{Gatti:1986cw} is valid for defined pulse shapes, i.e. for a strictly linear detector response. TES sensors have intrinsically a non-linear response due to natural shape of the superconducting transition curves~\cite{Strauss:2017cam}, leading to gradually increasing non-linearities at higher energy up to full saturation when reaching the normal-conduction state. The difference in heater and particle pulse shapes may differently affect the linearity of the OF response for both event classes. Only in the energy region of linear TES response the reconstruction of heater and particle pulses can be unambiguously matched by the OF. Energy reconstruction methods taking into account non-linearities, such as truncated template fit~\cite{Strauss:2017cam}, need to be further investigated for this data set. 

Secondly, potential hardware causes of heater pulse non-linearity are under investigation. The new data acquisition system (VDAQ02) used for operation and stabilization of this measurement campaign has not been validated regarding the linearity of heater pulse injected energy on the required level. 

On the analysis side, these findings have been confirmed by an independent analysis using different software and analyzers.

%% file: EnergyResolutionModel.tex
\section{Example of Applications: Modeling the Detector Energy Resolution}
\label{sec:eneryresolutionmodel}

Unlike conventional thermal phonon detection, TES-based macro-calorimeters operated at cryogenic temperatures capture prompt athermal phonon signals, significantly improving energy resolution for gram-scale detectors. This enhancement benefits rare event searches, necessitating both mass and low threshold, such as CRESST's low mass dark matter searches and the NUCLEUS experiment's measurements of coherent neutrino scattering at nuclear reactors.

For particle events within a crystal, the deposited energy is initially transferred to the lattice as high-frequency phonon modes, with phonon energies typically ranging from 0.5 to 1.5\,meV. These high-energy phonons quickly decay into athermal phonons that propagate ballistically through the crystal.
A fraction of the athermal phonons is absorbed by the TES or phonon collectors, where the phonon energy is downconverted and transmitted to the tungsten film with some losses \cite{Probst:1995hjq}. Measuring the overall athermal phonon collection efficiency is crucial for understanding and then optimizing detector performances.

As an attempt to further understand our detection system, and motivated by the precise XRF measurements that provide both energy and resolution as a function of energy, we developed a phenomenological model for the energy resolution of our detection system. This model incorporates three key components: a statistical term associated with athermal phonon statistics, a systematic term that scales with energy, and a constant term independent of energy representing the overall sensor noise. The statistical term, influenced by the athermal phonon collection efficiency, is parameterized by \(\eta\) and can be derived from the XRF resolution curve, which shows the relative resolution as a function of true energy. This approach enables an evaluation of the collection efficiency of our TES-based athermal phonon sensors.

The energy resolution model is given by:

\begin{equation}
\frac{\sigma(E)}{E} = \frac{\alpha}{\sqrt{N_{at}(E)}} \oplus \beta \oplus \frac{\sigma_0}{E}
\end{equation}

where:
the first term represents the statistical contribution from athermal phonon statistics, with \(\alpha = 1\) expected from Poisson statistics, \(\beta\) is a constant term representing systematic effects such as non-uniformities and dependence of collection efficiency on position of energy deposition in the target\footnotemark, the third term describes the noise provided by the baseline resolution \(\sigma_0\), and \(\oplus\) denotes the quadratic sum.

\footnotetext{In principle, insufficient temperature control and resulting detector instability could also contribute to this term, but this was found to be negligible in this measurement.}

The number of athermal phonons, \(N_{at}(E)\), is given by:

\begin{equation}
N_{at}(E) = E \times \frac{\eta}{e_{\text{ath}}}
\end{equation}

where:
\(E\) is the energy deposited (electron recoil in our case), \(\eta\) is the athermal phonon collection efficiency and \(e_{\text{ath}}\) is the mean energy of an athermal phonon, taken to be 1 meV in the following.
The value for the average athermal phonon energy is not known precisely and subject to an uncertainty of probably $\mathcal{O} (50\%)$ which directly propagates on the fitted value for $\eta$.

\begin{figure}[h!]
    \centering
    \includegraphics[width=0.45\textwidth]{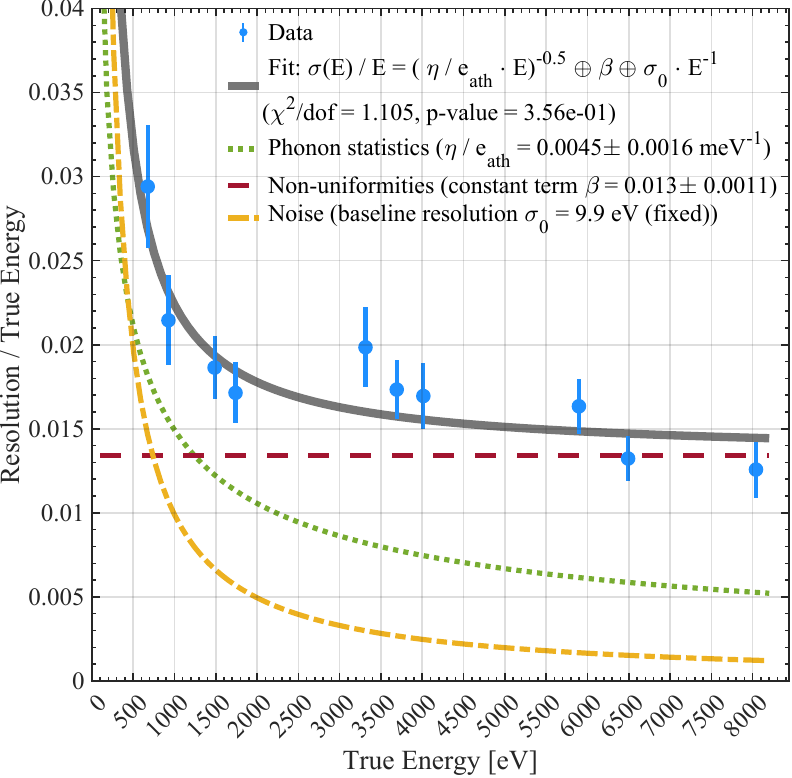}
    \caption{Fit of the energy resolution model to the XRF data. The plot shows the relative energy resolution \(\frac{\sigma(E)}{E}\) as a function of true energy \(E\). The best-fit components include the phonon statistics term, non-uniformities term, and baseline resolution term. The best-fit value for the ratio of athermal phonon collection efficiency over mean athermal phonon energy is $0.0045_{-0.0012}^{+0.0025}\frac{\text{1}}{\text{meV}}$, the non-uniformities constant term is \(\beta = 0.013 \pm 0.0011\), and the baseline resolution term is fixed to \(\sigma_0 = 9.9\) eV. The fit quality is indicated by \(\chi^2/\text{dof} = 1.105\) with a p-value of 0.36.}
    \label{fig:energy_resolution_model}
\end{figure}

The fit of the energy resolution model to the XRF data is shown in Figure \ref{fig:energy_resolution_model}. The figure plots the relative energy resolution \(\frac{\sigma(E)}{E}\) as a function of the true energy \(E\).
The best-fit value for the ratio $\eta/e_{\text{ath}}$ is  \(0.0045_{-0.0012}^{+0.0025} \frac{\text{1}}{\text{meV}}\) yielding \(\eta = 0.0045_{-0.0026}^{+0.0034}\). 
The asymmetric error is obtained by scanning over different values for $\eta/e_{\text{ath}}$, keeping $\eta/e_{\text{ath}}$ fixed in the fit and leaving only $\beta$ free. The constant term \(\beta\) is found to be $0.013 \pm 0.0011$, and the baseline resolution term \(\sigma_0\) is fixed to \(9.9 \text{ eV}\). The fit quality is indicated by \(\chi^2/\text{dof} = 1.105\), with a p-value of 0.36, attesting to a good description of the data.

The relative resolution is primarily influenced by one of three terms, depending on the energy. For energies greater than 2000 eV, the relative resolution is predominantly influenced by the constant term \(\beta\), which is 1.3\%. This term can be interpreted as a vertex-dependent broadening intrinsic to the detector system, representing a fixed fraction \(\beta\) of the true energy. This explains the relatively large resolutions (on the order of 100 eV) obtained for the \(^{55}\text{Mn}\) K$_\alpha$ and K$_\beta$ lines, for instance. At low energies (less than 500 eV), the resolution is dominated by the baseline resolution \(\sigma_0\) of the detector, corresponding to the intrinsic noise of the data stream. This resolution is measured by using noise samples and simulated pulse reconstruction after applying the XRF calibration. Consequently, it is a fixed parameter in our fit. For intermediate energies (\(\text{500 eV} < E < \text{2000 eV}\)), the relative resolution is influenced by the baseline resolution \(\sigma_0\), the constant term \(\beta\), and the athermal phonon collection efficiency \(\eta\). For \(\eta\) values below 1\%, this term can dominate part of the energy range. Therefore, our model provides a novel approach to measuring the phonon collection efficiency, as reported for the first time in this publication.

We find a 95\% upper limit on the ratio $\eta/e_{\text{ath}}$ that corresponds to a phonon collection efficiency of 1.45\% for the assumed mean phonon energy of \SI{1}{meV}.
Our results thus indicate a collection efficiency ranging from a fraction of a percent to about one percent. This number can be independently cross-checked with estimates obtained from a model of heat capacities and thermal conductances in the detector and the TES signal response.

In conclusion, this study aims to demonstrate a proof-of-concept for assessing the athermal phonon efficiency. More precise data would be required to achieve a more accurate estimation of the collection efficiency. Moreover, a careful assessment of the systematic uncertainties in the resolution $\sigma(E)$ of each peak would be required, which is beyond the scope of this publication.

%% file: Conclusion.tex
\section{Conclusion and Outlook}
\label{sec:Outlook}

In this article, we introduced several advancements in the calibration and analysis of cryogenic macro-calorimeters achieving sub-100 eV thresholds, essential for light DM searches and neutrino studies based on CEvNS. We detailed the development of a novel XRF calibration source, utilizing an initial $^{55}$Fe source to irradiate a two-staged target arrangement, emitting a spectrum of calibration lines from 677\,eV to 6.5\,keV, enabling a comprehensive investigation of the detector response. We described the 18.7 days of data acquisition with the XRF source using the NUCLEUS CaWO$_4$ detector prototype equipped with TES. 

The three most prominent lines in the resulting cryodetector spectrum are Mn K$_\alpha$, Al K$_\alpha$, and Ca K$_\alpha$, which are all suppressed by one or two elastic scatterings. This suggests that the figure-of-merit of our source could be improved further by reducing the probability of elastic scattering onto the detector, e.g. by changing the angle of the targets away from $45^\circ$. There is a trade-off between this reduction and the loss of solid angle for XRF photons between the stages, which can be further studied in future iterations of the source configuration. In addition, there could be further developments by miniaturizing the source setup or using a \(^{55}\)Fe source with higher activity.

The advanced CryoLab analysis tool was introduced, featuring sophisticated methods for data processing, calibration, and high-level analysis to ensure robust, efficient, and scalable data management with the option to work on real and synthetic data in parallel.
We presented the analysis of the XRF data, including the derivation of the calibration curve and spectrum extraction, successfully extending the energy range of XRF calibration into the sub-keV region relevant for light DM and CEvNS studies~\cite{Strauss:2017cam, Angloher:2019flc, CRESST:2019jnq}. This extension enables an assessment of detector linearity within this critical energy range.
The linear Mn K$_\alpha$ calibration was found to be strongly disfavored with respect to the non-linear model.
Additionally, we explored the potential of studying detector physics through energy resolution as a function of energy. 

It is important to note that the XRF measurement presented in this work, and also the in-situ calibration strategy of the NUCLEUS experiment using LED bursts \cite{Cardani:2021iff}, involve the calibration of the energy scale assuming electron-like recoils. For CEvNS and Dark Matter experiments, it is also crucial to calibrate nuclear recoil signals, as done by the CRAB collaboration \cite{PhysRevLett.130.211802}. As a next step, a simultaneous calibration of both electron (XRF) and nuclear recoils in the sub-keV range would be highly valuable for achieving more robust and final calibration results.

%% file: xrf.bbl
\providecommand{\noopsort}[1]{}\providecommand{\singleletter}[1]{#1}%

%% file: Appendix.tex
\section*{Appendix}

\begin{figure}[ht!]
    \centering
    \includegraphics[width=0.9\linewidth]{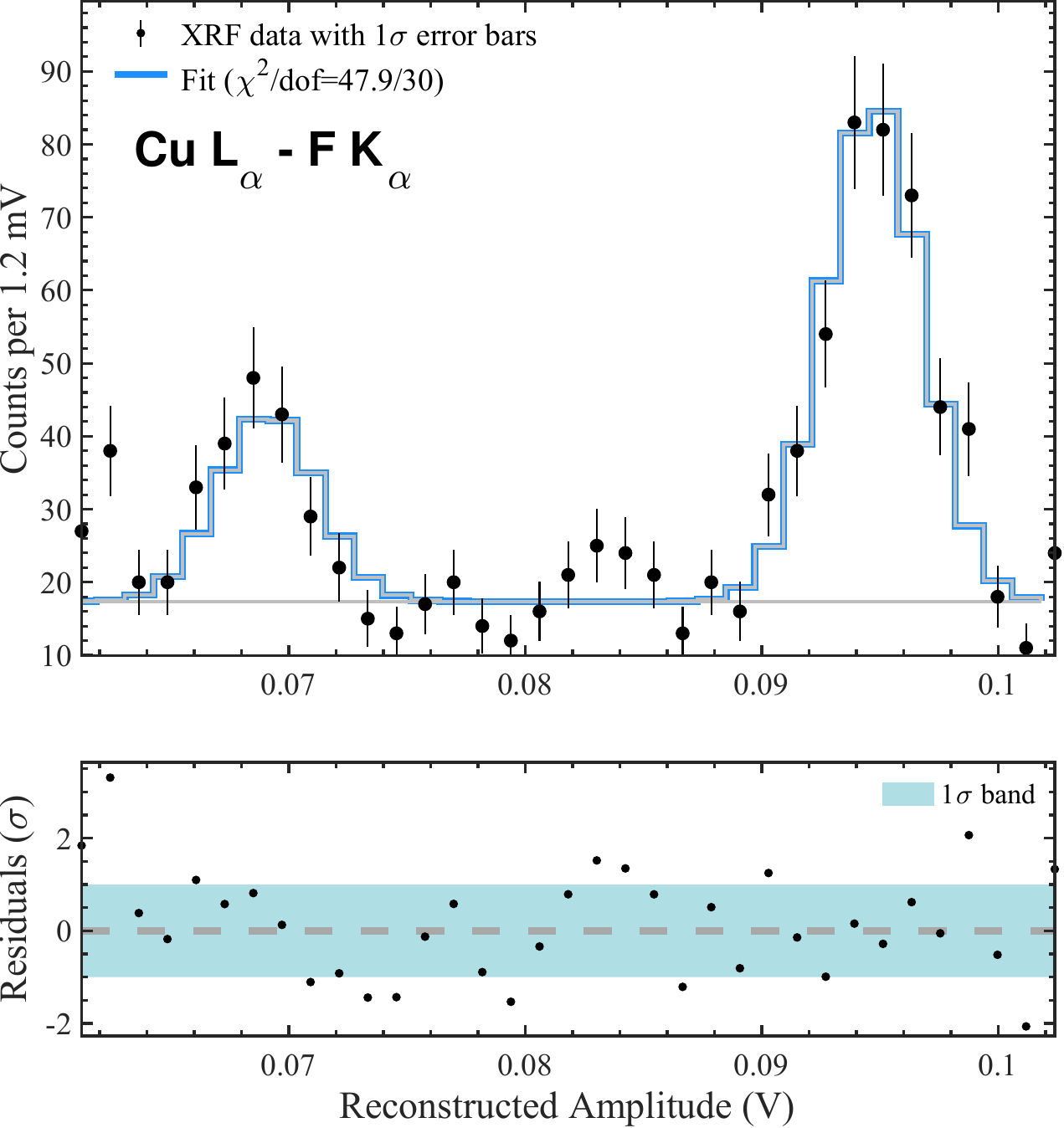} 
    \caption{Counts as a function of reconstructed amplitude and display of the F K$_\alpha$ (676.8 eV) and Cu L$_\alpha$ (927.7 eV) line fits using Gaussian models. The corresponding fit values are quoted in Table \ref{tab:xray_results}.}
    \label{fig:peakfit_FCu}
\end{figure}
\begin{figure}[ht!]
    \centering
    \includegraphics[width=0.9\linewidth]{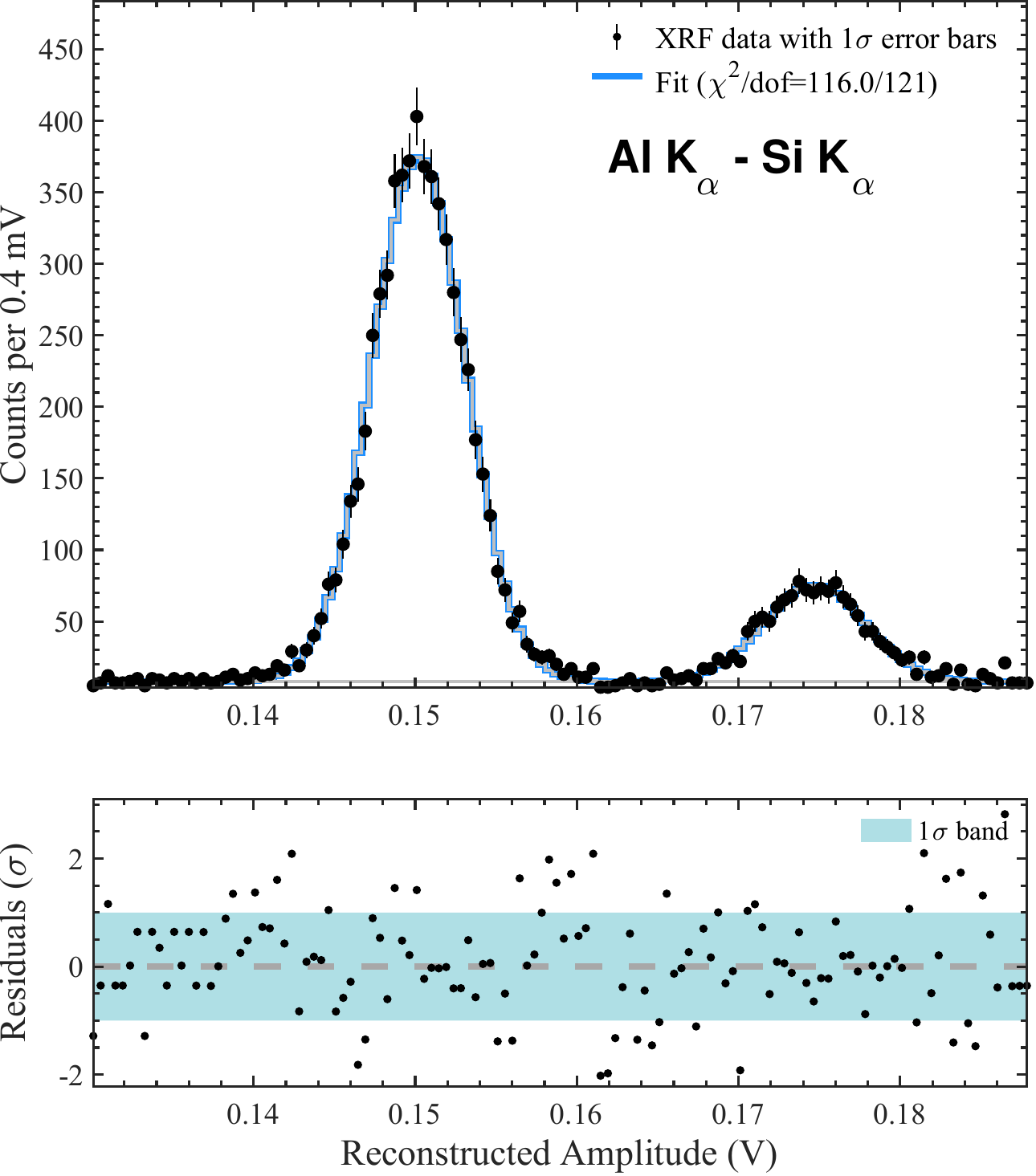}
    \caption{Counts as a function of reconstructed amplitude and display of the Al K$_\alpha$ (1486.4 eV) and Si K$_\alpha$ (1739.6 eV) line fits using Gaussian models. The corresponding fit values are quoted in Table \ref{tab:xray_results}.}
    \label{fig:peakfit_AlSi}
\end{figure}
\begin{figure}[ht!]
    \centering
    \includegraphics[width=0.9\linewidth]{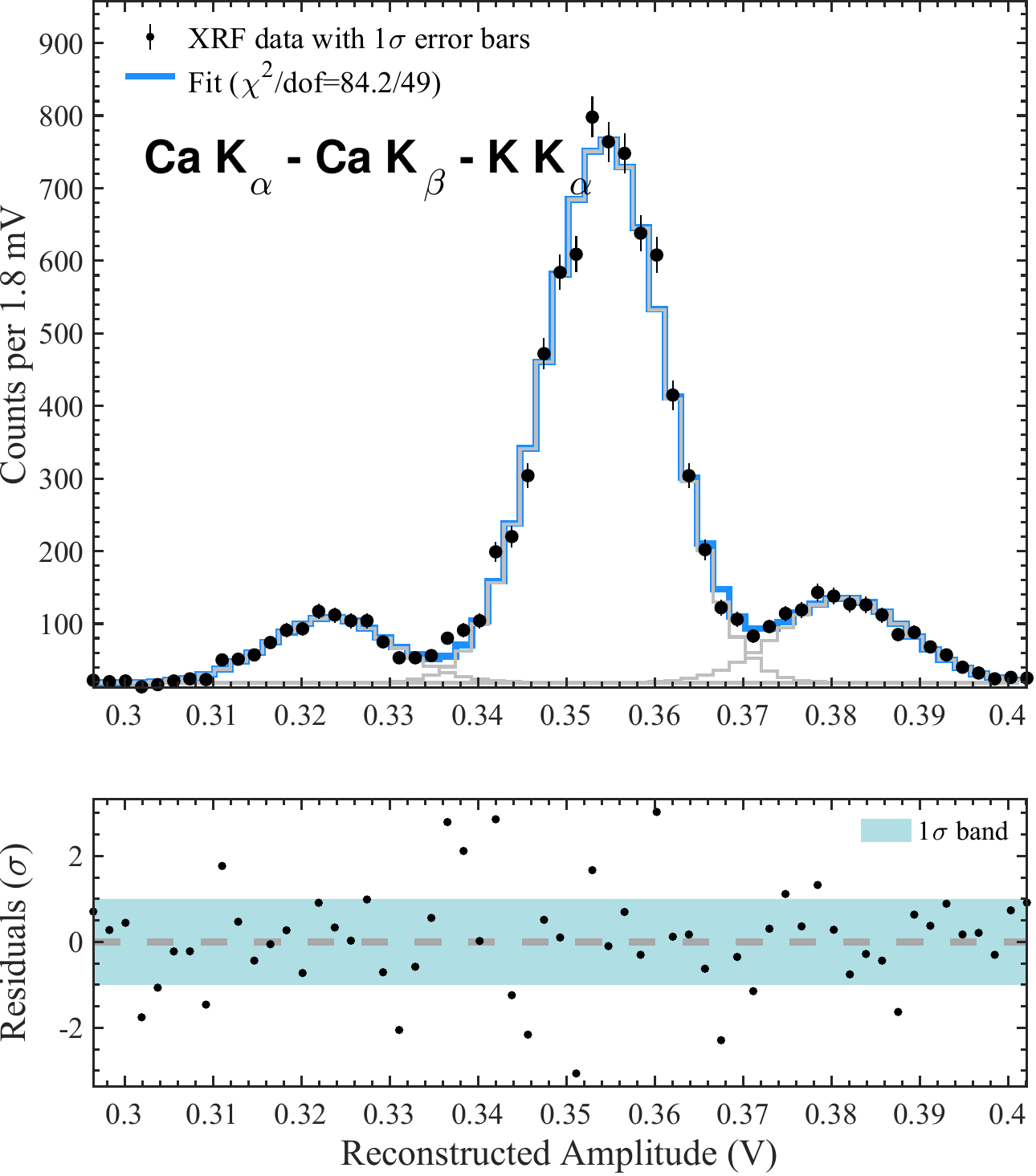} 
    \caption{Counts as a function of reconstructed amplitude and display of the K K$_\alpha$ (3312.9 eV), Ca K$_\alpha$ (3691.1 eV) and Ca K$_\beta$ (4013.1 eV) line fits using Gaussian models. The corresponding fit values are quoted in Table \ref{tab:xray_results}.}
    \label{fig:peakfit_KCa}
\end{figure}
\begin{figure}[ht!]
    \centering
    \includegraphics[width=0.9\linewidth]{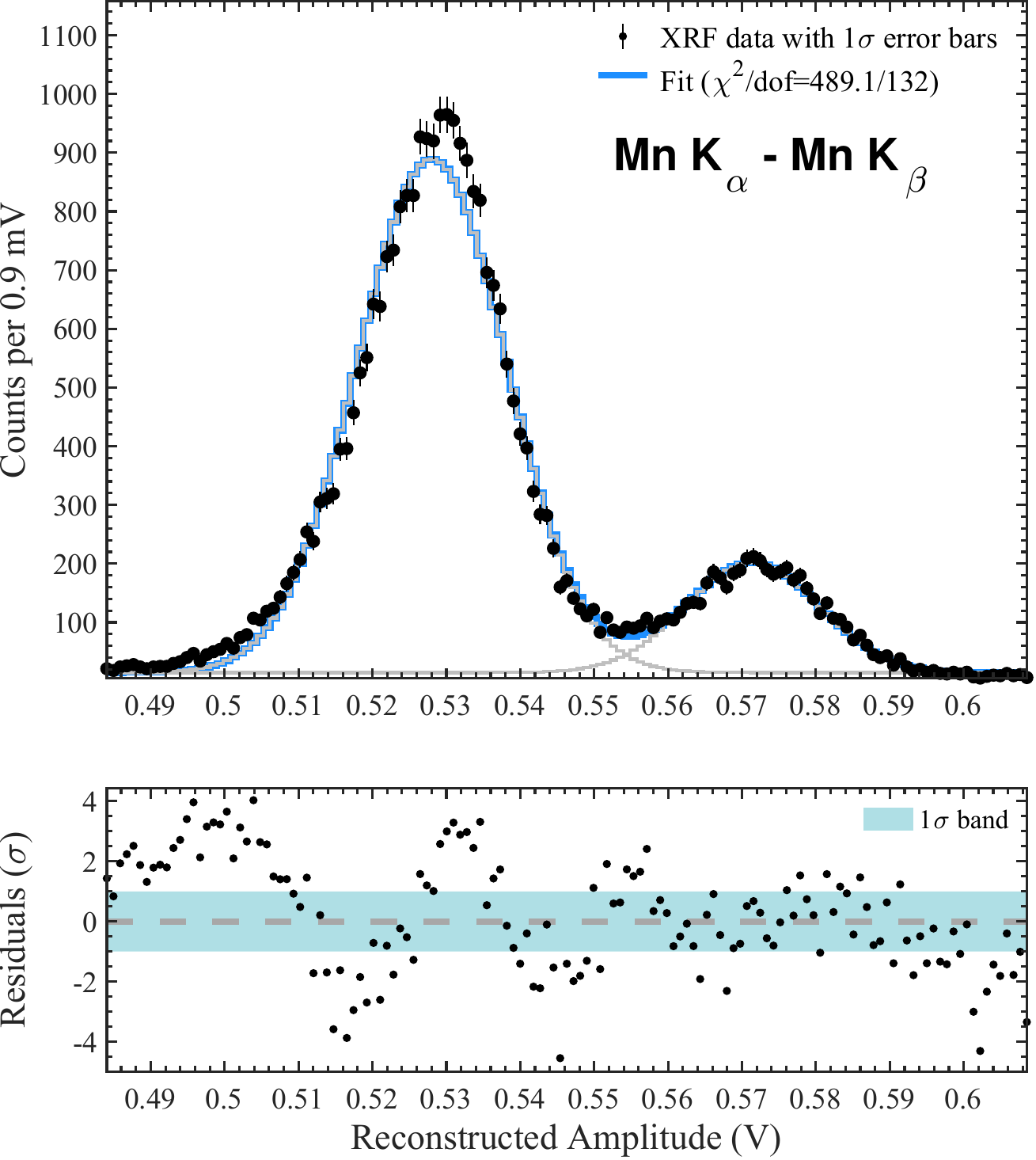} 
    \caption{Counts as a function of reconstructed amplitude and display of the Mn K$_\alpha$ (5896.5 eV) and Mn K$_\beta$ (6491.8 eV) line fits using Gaussian models. The corresponding fit values are quoted in Table \ref{tab:xray_results}.}
    \label{fig:peakfit_Mn}
\end{figure}
\begin{figure}[ht!]
    \centering
    \includegraphics[width=0.9\linewidth]{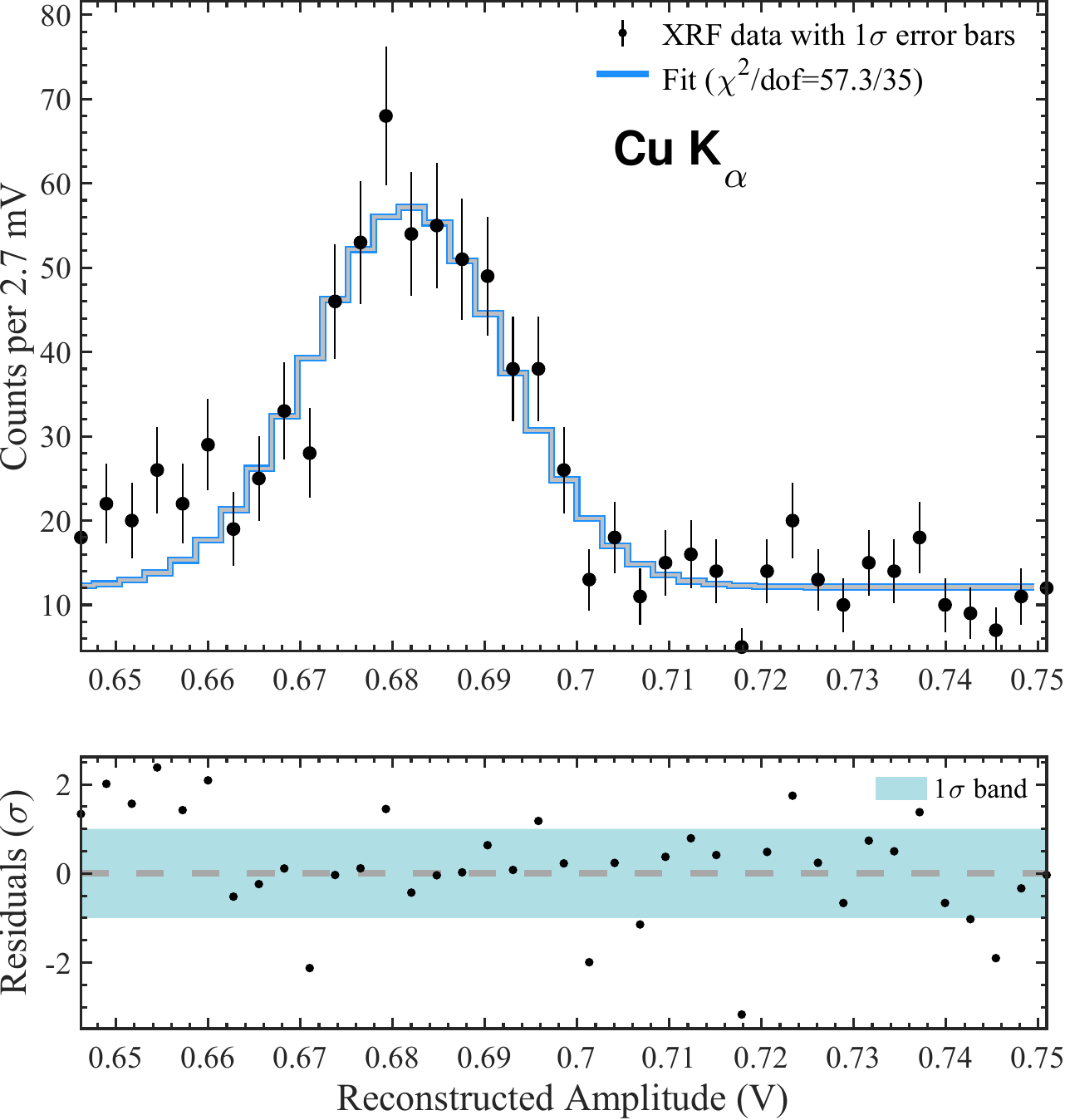} 
    \caption{Counts as a function of reconstructed energy and display of the Cu K$_\alpha$ (8039.7 eV) line fit using a Gaussian model. The corresponding fit values are quoted in Table \ref{tab:xray_results}.}
    \label{fig:peakfit_Cu8keV}
\end{figure}